\documentclass[a4paper,fleqn]{cas-dc}

\usepackage[numbers]{natbib}
\usepackage[utf8]{inputenc}
\DeclareUnicodeCharacter{22EF}{\dots}
\usepackage{soul}
\usepackage{siunitx}

\usepackage{soul}

\def\tsc#1{\csdef{#1}{\textsc{\lowercase{#1}}\xspace}}
\tsc{WGM}
\tsc{QE}
\tsc{EP}
\tsc{PMS}
\tsc{BEC}
\tsc{DE}

\begin{document}
\let\WriteBookmarks\relax
\def\floatpagepagefraction{1}
\def\textpagefraction{.001}
\shorttitle{Applied Surface Science}
\shortauthors{Martins et~al.}

\title [mode = title]{HOP-graphene: A high-capacity anode for Li/Na-ion batteries unveiled by first-principles calculations}
           
\author[1]{ Nicolas F. Martins}
\affiliation[1]{
organization={Modeling and Molecular Simulation Group},
addressline={São Paulo State University (UNESP), School of Sciences}, 
city={Bauru},
postcode={17033-360}, 
state={SP},
country={Brazil}}
\cormark[1]
\cortext[cor1]{Principal corresponding author}
\ead{nicolas.ferreira@unesp.br}
\credit{Conceptualization of this study, Methodology, Investigation, Formal analysis, Writing -- review \& editing, Writing -- original draft}

\author[1]{José A. S. Laranjeira}
\credit{Conceptualization of this study, Methodology, Investigation, Formal analysis, Writing -- review \& editing, Writing -- original draft}

\author[2]{Kleuton A. L. Lima}
\affiliation[2]{
organization={Department of Applied Physics and Center for Computational Engineering and Sciences},
addressline={State University of Campinas}, 
city={Campinas},
postcode={13083-859}, 
state={SP},
country={Brazil}}
\credit{Conceptualization of this study, Methodology, Review and editing, Investigation, Formal analysis, Writing -- review \& editing, Writing -- original draft}

\author[3]{{Luis A. Cabral}}
\credit{Investigation, Formal analysis, Resources, Writing -- review \& editing}
\affiliation[4]{organization={Department of Physics and Meteorology},
addressline={São Paulo State University (UNESP), School of Sciences}, 
city={Bauru},
postcode={17033-360}, 
state={SP},
country={Brazil}}

\author[4,5]{L.A. Ribeiro Junior}
\affiliation[4]{
organization={Institute of Physics},
addressline={University of Brasília}, 
city={Brasília },
postcode={70910‑900}, 
state={DF},
country={Brazil}}
\affiliation[5]{
organization={Computational Materials Laboratory, LCCMat, Institute of Physics},
addressline={University of Brasília}, 
city={Brasília },
postcode={70910‑900}, 
state={DF},
country={Brazil}}
\credit{Conceptualization of this study, Methodology, Review and editing, Investigation, Formal analysis, Writing -- review \& editing, Writing -- original draft}

\author[1]{Julio R. Sambrano}
\credit{Formal analysis, Resources, Writing -- review \& editing}
\cormark[2]
\cortext[cor2]{corresponding author}

\begin{abstract}
The growing demand for efficient energy storage has driven the search for advanced anode materials for lithium- and sodium-ion batteries (LIBs and SIBs). In this context, we report the application of HOP-graphene (a 5-6-8-membered 2D carbon framework) as a high-performance anode material for LIBs and SIBs using density functional theory simulations. Diffusion studies reveal low energy barriers of 0.70 eV for Li and 0.39 eV for Na, indicating superior mobility at room temperature compared to other carbon allotropes, like graphite. Full lithiation and sodiation accommodate 24 Li and 22 Na atoms, respectively, delivering outstanding theoretical capacities of 1338 mAh/g (Li) and 1227 mAh/g (Na). Bader charge analysis and charge density difference maps confirm substantial electron transfer from the alkali metals to the substrate. Average open-circuit voltages of 0.42 V (Li) and 0.33 V (Na) suggest favorable electrochemical performance. HOP-graphene also demonstrates excellent mechanical strength. These findings position HOP-graphene as a promising candidate for next-generation LIB and SIB anodes.
\end{abstract}


\maketitle
\section{Introduction}

The ability of carbon to form bonds with diverse hybridizations, combined with the robustness of its covalent interactions, underpins its exceptional chemical versatility~\cite{kong2019sp2, vejpravova2021mixed}. Since the successful synthesis of graphene in 2004~\cite{geim2007rise}, the pursuit of new two-dimensional (2D) materials with atomic thickness has become a major focus of research~\cite{nawaz2025flatland,ares2022recent}. This effort is fueled by the remarkable adaptability in synthesizing novel carbon-based materials, achieved by carefully selecting precursor units~\cite{fan2019nanoribbons, girao2023classification}. Several such materials, graphdiyne~\cite{gao2019graphdiyne}, triphenylene-graphdiyne~\cite{matsuoka2018expansion}, biphenylene~\cite{fan2021biphenylene}, and graphenylene~\cite{du2017new}, have been successfully synthesized, while many others have been proposed through computational simulations \cite{enyashin2011graphene,shi2021high}.

For instance, Zhang \textit{et al.}~\cite{zhang2024prediction} employed azulenoid kekulenes (AK) as precursor agents to predict various $\pi$-conjugated 2D materials featuring porous and non-porous architectures, some exhibiting semi-metallic or semiconducting characteristics. More recently, density functional theory (DFT) simulations have been used to evaluate 30 distinct semiconducting 2D carbon allotropes, demonstrating their significant promise for photovoltaic applications~\cite{cavalheiro2024can}. High-throughput systematic studies have also been reported on the prediction of new materials~\cite{xie2020general,al2021high,blatov2021high}. Among the most notable predicted monolayers are graphenyldiene~\cite{laranjeira2024graphenyldiene}, irida-graphene~\cite{junior2023irida}, beta-Naphthyne~\cite{laranjeira20252d}, and Pan-C$_2(n+1)$~\cite{liu2025pan}.

Due to their ring-based architectures that differ from conventional graphene sheets, numerous 2D carbon allotropes have been explored for applications in energy storage systems~\cite{rajkamal2019carbon,tiwari2016magical,folorunso2024heteroatom}. Their unique geometries are particularly favorable for metal-ion diffusion, offering more accessible migration pathways than commercial anodes such as graphite~\cite{chae2020integration}. For instance, Yu~\cite{yu2013graphenylene} investigated the graphenylene lattice and identified lithium interlayer diffusion barriers ranging from 0.25 to 0.57 eV, highlighting the role of 4-6-12-membered rings in facilitating ion mobility. Furthermore, Ferguson et al.~\cite{ferguson2017biphenylene} demonstrated, through theoretical modeling, that lithium migration in biphenylene exhibits energy barriers below (\(< 0.45\) eV), comparable to those in graphene (\(< 0.46\) eV)~\cite{liu2017all}.

In addition to lithium-ion batteries (LIBs), alternative technologies have been actively pursued to address concerns related to energy supply and cost~\cite{fan2020sustainable}. Among various energy storage technologies, sodium-ion batteries (SIBs) have attracted considerable attention due to the abundance and affordability of this alkali metal \cite{vaalma2018cost}. However, while lithium can deliver a capacity of 372 mAh/g on graphite anodes~\cite{armand2008building}, sodium storage on the same substrate is severely limited to just 35 mAh/g~\cite{wang2016emerging}. This discrepancy has prompted the search for alternative anode materials better suited to sodium storage~\cite{deshmukh2025sodium,perveen2020prospects}. Similarly, numerous reports have positioned 2D carbon allotropes as viable alternatives for non-LIB sources. Notable examples include successful theoretical studies of Na storage in graphenylene~\cite{fabris2021promising}, biphenylene~\cite{chen2023biphenylene}, irida-graphene~\cite{martins2024irida}, graphene+\cite{surila2025doping}, and T-graphene~\cite{yadav2022si}.

In the ongoing search for advanced anode materials for both LIBs and SIBs, we investigate HOP-graphene as a promising alternative for energy storage via first-principles simulations. The HOP-graphene monolayer was first predicted by Mandal et al.~\cite{mandal2013theoretical} in 2013 as a new 2D carbon allotrope composed of 4-, 5-, and 8-membered rings, which exhibit intrinsic metallicity and dynamic stability. Since its proposal, several studies have explored the influence of anisotropy, strain rate, size, and defects on the electronic and mechanical properties of HOP-graphene~\cite{sui2017morphology, ren2025molecular, peng2024atomistic}. However, there remains a significant gap in research focused on understanding the practical applications of this new carbon monolayer.

\begin{figure*} [!ht]
    \centering
    \includegraphics[width=1\linewidth]{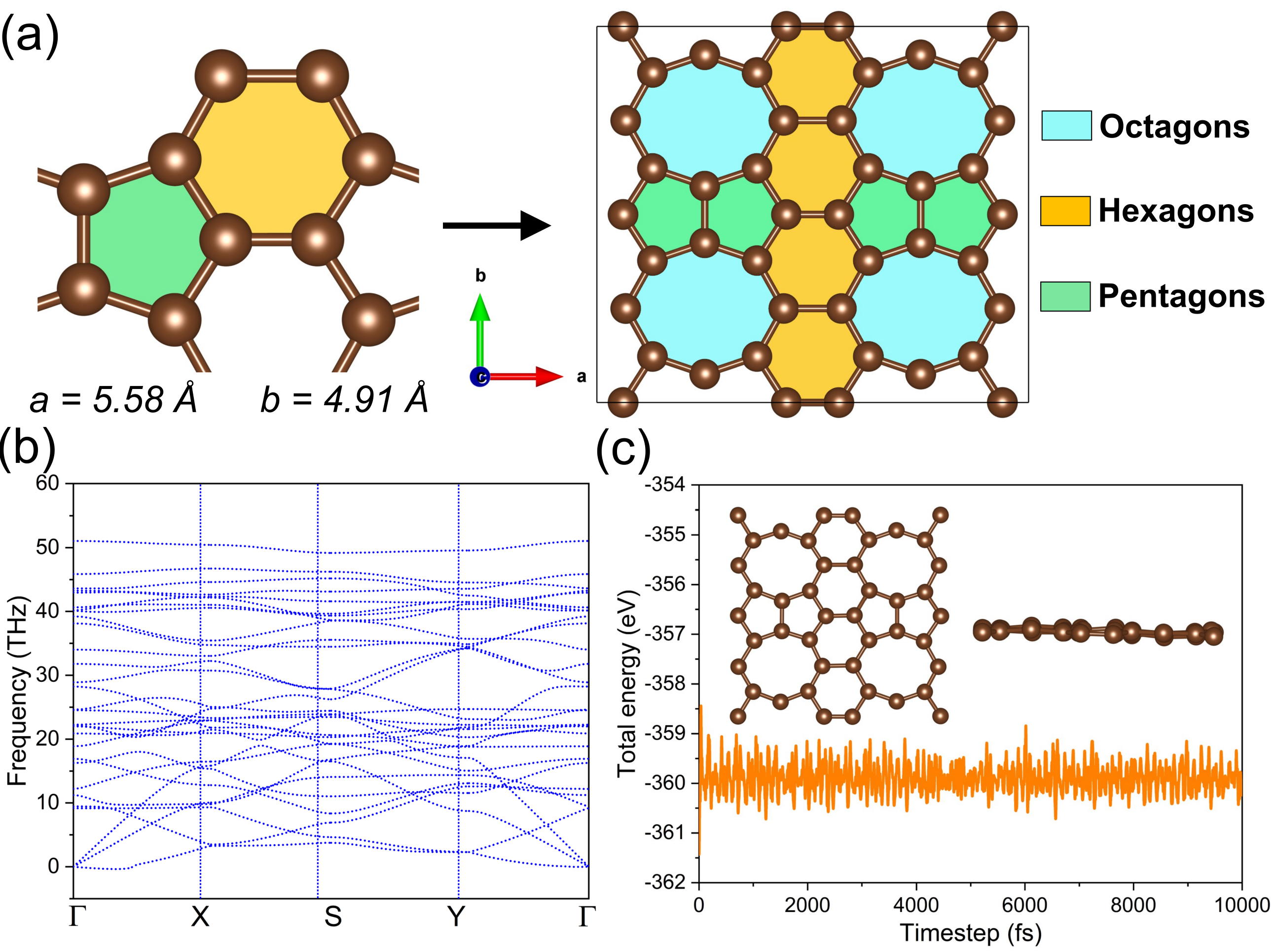}
    \caption{Structural and stability characterization of HOP-graphene. (a) Unit cell and extended supercell of HOP-graphene, highlighting the periodic arrangement of octagons (cyan), hexagons (yellow), and pentagons (green), with optimized lattice constants. (b) Phonon dispersion spectrum along high-symmetry paths of the Brillouin zone. (c) The total energy profile from AIMD simulations at 300 K over 10 ps. The inset shows top and side views of the structure at the end of the simulation.}
    \label{fig:structure}
\end{figure*}

In this study, we employ DFT simulations to analyze the adsorption of Li and Na on the HOP-graphene structure. Several key parameters were investigated, including adsorption energies, charge density difference (CDD) maps, Bader charge analysis, metal diffusion, and the open-circuit voltage (OCV) profile. Additionally, \textit{ab initio} molecular dynamics (AIMD) simulations were performed to evaluate the thermal stability and practical viability of HOP-graphene as an anode material for both lithium- and sodium-ion batteries.

\section{Computational Setup}

To characterize the HOP-graphene and assess its potential as an anode material, we performed DFT~\cite{becke2014perspective} and AIMD simulations~\cite{marx2000ab}. These calculations utilized the generalized gradient approximation (GGA) as formulated by Perdew, Burke, and Ernzerhof (PBE)~\cite{PhysRevLett.77.3865}. The projector augmented wave (PAW) method~\cite{PhysRevB.50.17953} was also employed, and both approaches were implemented within the Vienna \textit{ab initio} Simulation Package (VASP)]~\cite{kresse1993ab,kresse1996efficient}. The electronic wavefunctions were expanded using a plane-wave basis set with a kinetic energy cutoff of 520 eV. A vacuum spacing of 15 \r{A} was introduced along the $z$-axis to mitigate interactions between periodic images. 

Structural optimization and projected density of states (PDOS) calculations were performed using $\Gamma$-centered k-point meshes with grid sizes of $6 \times 6 \times 1$ and $9 \times 9 \times 1$, respectively. Dispersion interactions were accounted for using the DFT-D2 correction scheme proposed by Grimme~\cite{grimme2006semiempirical,moellmann2014dft}. Structural relaxation was carried out using the conjugate gradient method, with convergence criteria requiring that the total energy variations for atomic and lattice parameters remained below $1 \times 10^{-5}$ eV, and the Hellmann-Feynman forces on individual atoms did not exceed 0.01 eV/\r{A}. AIMD simulations were performed to assess the thermodynamic stability of both Li and Na storage on HOP-graphene. In these simulations, a time step of 0.5 fs was employed, with a total simulation time of 5 ps using the NVT ensemble within the Nos\'e-Hoover thermostat~\cite{hoover1985canonical}, at 300 K. Additionally, the Bader charge decomposition method was utilized to analyze the charge transfer mechanism \cite{henkelman2006fast}. The nudged elastic band (NEB) method was also employed to investigate the Li and Na diffusion over the HOP-graphene surface.

\section{Results and Discussion}

\subsection{Structural, electronic, and mechanical properties of HOP-graphene}

HOP-graphene has a rectangular network with lattice parameters \(a = 5.58\) \AA\ and \(b = 4.91\) \AA. Its unit cell is composed of 10 atoms, which, when expanded, generate a multi-member architecture of full octagons, hexagons, and pentagons, as depicted in Fig.~\ref{fig:structure}(a). This shape leads to six distinct bond lengths, ranging from 1.36 to 1.50 \AA, in agreement with reference~\cite{mandal2013theoretical}. Phonon dispersion curves were assessed to verify the dynamic stability of HOP-graphene (see Fig.~\ref{fig:structure}(b). Generally, the frequencies are positive along the high-symmetry paths of the Brillouin zone, with three acoustic modes below 15 THz. However, minor imaginary frequencies are observed near the \(\Gamma\) point, approximately at a frequency of -0.5 THz, which may result from different technical aspects, such as supercell size or pseudopotentials. As noted by Wang et al.~\cite{wang2022high}, the stability of a 2D material can be supported for frequency outputs up to -2 THz, within the range verified by the HOP-graphene monolayer. Fig.~\ref{fig:structure}(c) shows the energetic profile of the AIMD simulations performed over 10 ps at 300 K for pristine HOP-graphene. Almost negligible energy fluctuations and the structural integrity of the monolayer (see the final AIMD snapshots in the inset) attest to the thermal stability of HOP-graphene.

\begin{figure} [!ht]
    \centering
    \includegraphics[width=\linewidth]{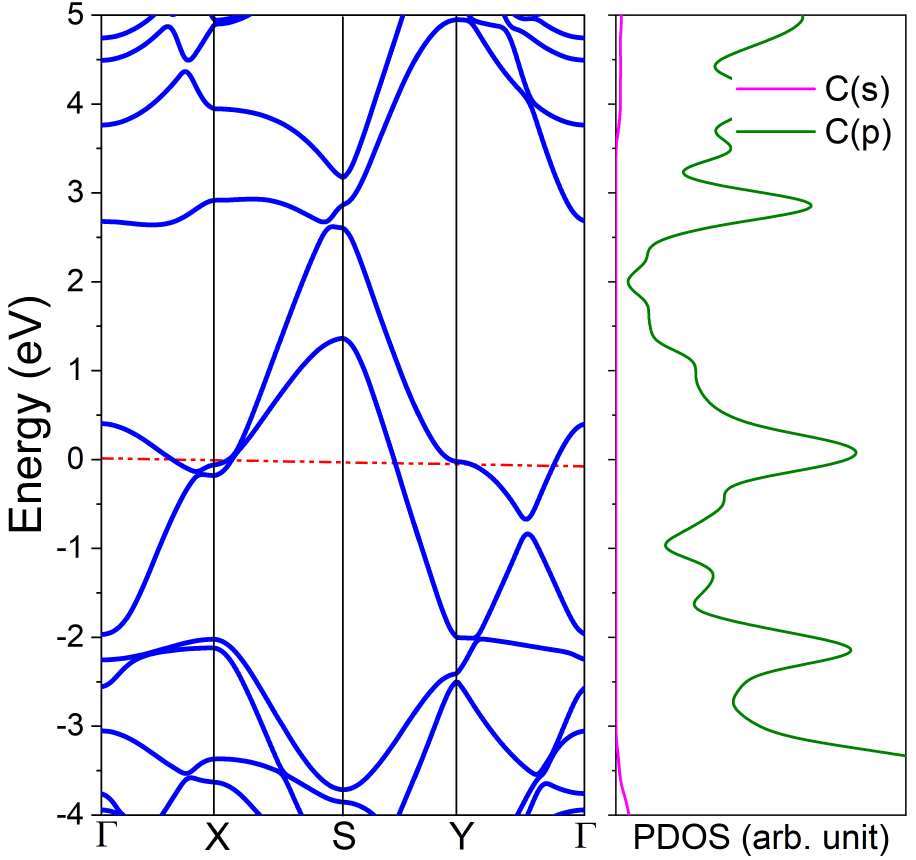}
    \caption{Band structure and projected density of states (PDOS) for pristine HOP-graphene. Fermi level is fixed at zero eV for better illustration.}
    \label{fig:PDOS pristine}
\end{figure}

The electronic response of HOP-graphene was also evaluated by analyzing its band structure and projected density of states (PDOS), as shown in Fig.~\ref{fig:PDOS pristine}. The results reveal the strong metallic character of this carbon monolayer, with several bands crossing the Fermi level, which is set at zero. The overlapping band regions show significant contributions from the C($p$) orbitals. These findings highlight the potential of HOP-graphene to function as an effective electrode material, as intrinsic conductivity is crucial for fast charge/discharge processes and reliable battery performance.

\begin{figure*}
    \centering
    \includegraphics[width=0.75\linewidth]{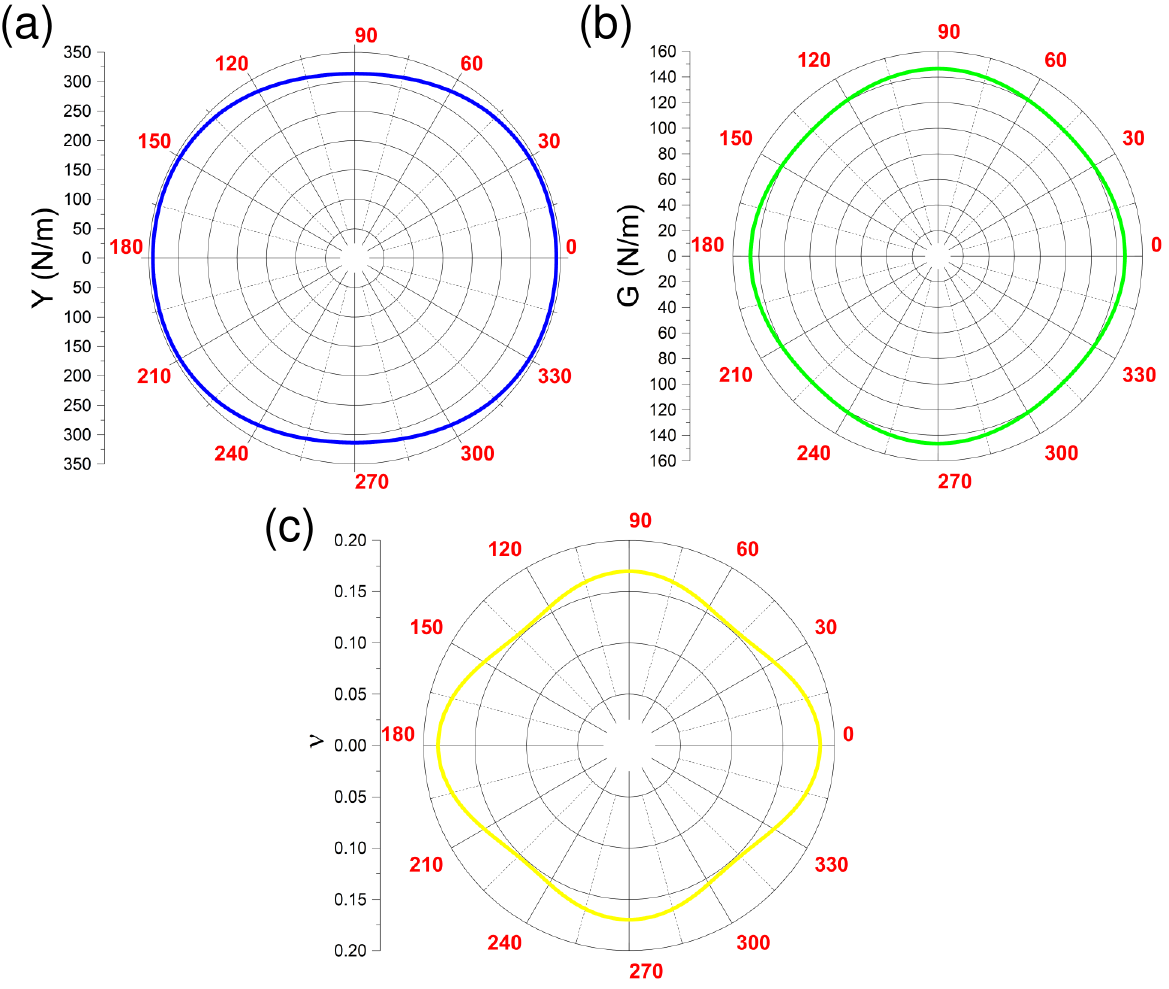}
    \caption{Polar diagrams representing (a) Young Modulus (\( Y \)), (b) Shear Modulus (\( G \)), and (c) Poisson's ratio (\( \nu \)) of HOP-graphene. }
    \label{fig:mech}
\end{figure*}

The mechanical properties of HOP-graphene were analyzed, as illustrated by the representations in the plane of \( Y \), \( G \), and \( \nu \) in Fig.~\ref{fig:mech}. Young's modulus (\( Y \)) has a maximum (minimum) value of 313.45 (347.67) N/m, with a small anisotropy ratio. Similarly, for the shear modulus (\( G \)), the highest and lowest values are 139.14 N/m and 146.35 N/m, respectively. The Poisson's ratio (\( \nu \)) also shows a negligible variation, ranging from a maximum of 0.15 to a minimum of 0.19, resulting in an exceptionally high anisotropy. This way, HOP-graphene can be classified as an almost isotropic material concerning its mechanical properties. The mechanical stability of HOP-graphene was evaluated in terms of its elastic constants, where \( C_{11} = 354.33 \) N/m, \( C_{22} = 323.65 \) N/m, \( C_{12} = 60.12 \) N/m, and \( C_{66} = 146.35 \) N/m. A mechanically stable 2D material with rectangular symmetry must satisfy the conditions \( C_{11} > 0 \), \( C_{66} > 0 \), and \( C_{11} C_{22} > C_{12}^2 \), according to the Born–Huang mechanical stability criteria for orthorhombic crystals \cite{PhysRevB.90.224104,doi:10.1021/acs.jpcc.9b09593}. HOP-graphene meets all these conditions.

\subsection{Binding of Li and Na atoms on HOP-graphene}

\begin{figure} [!ht]
    \centering
    \includegraphics[width=1.00\linewidth]{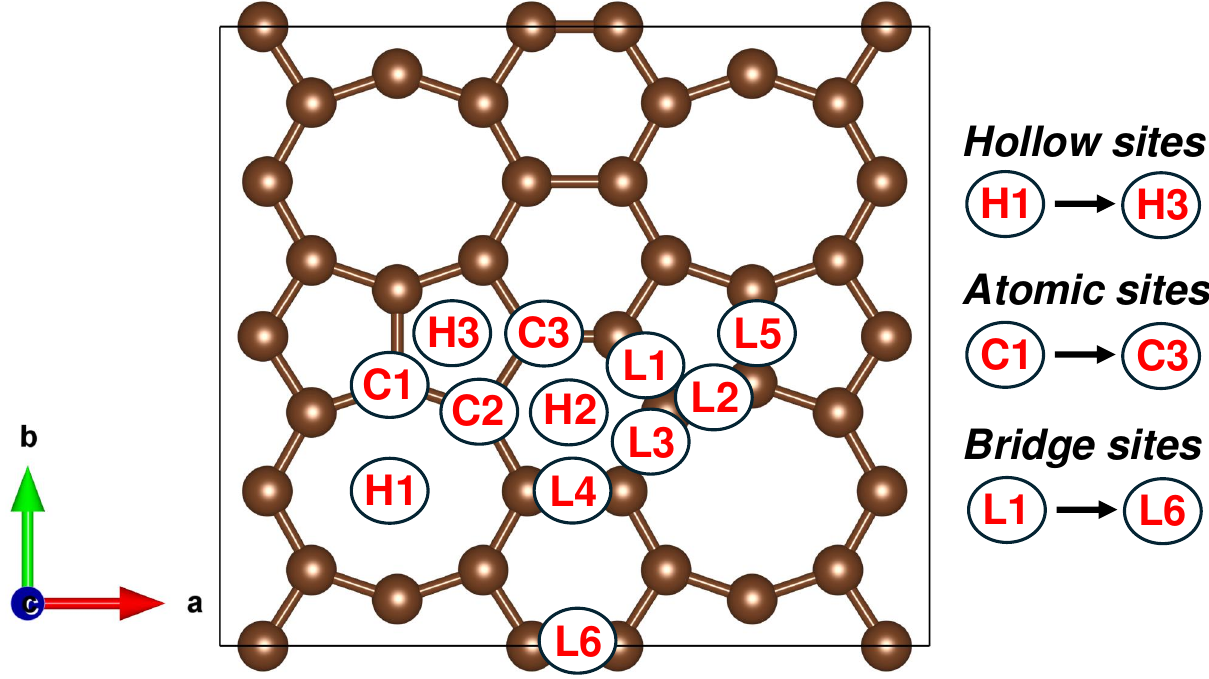}
    \caption{Different Li and Na adsorption sites over HOP-graphene monolayer.}
    \label{fig:sites}
\end{figure}

The capacity of the HOP-graphene monolayer to retain Li/Na ions is investigated by evaluating the adsorption energies ($E_{\text{ads}}$) through single-atom adsorption analysis. The $E_{\text{ads}}$ was calculated using the following expression:
\begin{equation}
  E_{\text{ads}} = E_{\text{(HOP + Li(Na))}} - E_{\text{(HOP)}} - E_{\text{(Li(Na))}}
\end{equation}

\noindent where $E_{\text{(HOP + Li(Na))}}$ represents the total energy of the Li(Na) adsorption on HOP-graphene, $E_{\text{HOP}}$ is the energy of the pristine HOP-graphene substrate, and $E_{\text{Li(Na)}}$ corresponds to the energy of an isolated Li or Na atom. A more negative value of E$_\text{ads}$ indicates a more thermodynamically favorable (i.e., stable) adsorption process. Fig.~\ref{fig:sites} shows the 12 available sites on HOP-graphene for alkali metal atom adsorption, classified as hollow sites (H1 to H3), atomic sites (C1 to C3), and bridge sites (L1 to L6).

\begin{figure} [!ht]
  \centering
    \includegraphics[width=\linewidth]{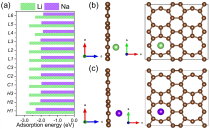}
    \caption{(a) Adsorption energies results for single Li(Na) atom adsorption on each available site of HOP-graphene monolayer. Side and top perspectives for (b) Li and (c) Na adsorption at the most stable configuration (H1 site)}
    \label{fig:Eads single}
\end{figure}

As expected, Li exhibits a stronger affinity for the HOP-graphene monolayer than Na, with adsorption energies ranging from -2.36 eV (L6) to -2.92 eV (H1). In comparison, Na adsorption energies lie between -1.87 eV (L6) and -2.32 eV (H1), indicating that the octagonal hollow site (H1) is the most favorable location for adsorbing both alkali metals. The moderate strength of these adsorption energies helps prevent the aggregation of Li or Na atoms into clusters during the charge/discharge cycles, thus enhancing operational safety. Figure~\ref{fig:Eads single}(a) presents the adsorption energy profiles for all evaluated sites, while Figures~\ref{fig:sites}(b) and \ref{fig:sites}(b) depict the most stable adsorption configurations for Li and Na, respectively.

Calculations show a higher binding strength for HOP-graphene compared to other reported 2D materials, such as irida-graphene (-2.58/-1.82~eV)~\cite{zhang2024li, martins2024irida}, B-doped biphenylene (-0.51/-1.14~eV)~\cite{fardi2024dft}, and T-graphene (-1.16/-1.13~eV)~\cite{zhang2020record}. These findings suggest that both Li and Na exhibit energetically stable adsorption on the HOP-graphene surface, a feature that may also benefit other storage applications, such as hydrogen.

\begin{figure} [!ht]
    \centering
    \includegraphics[width=1\linewidth]{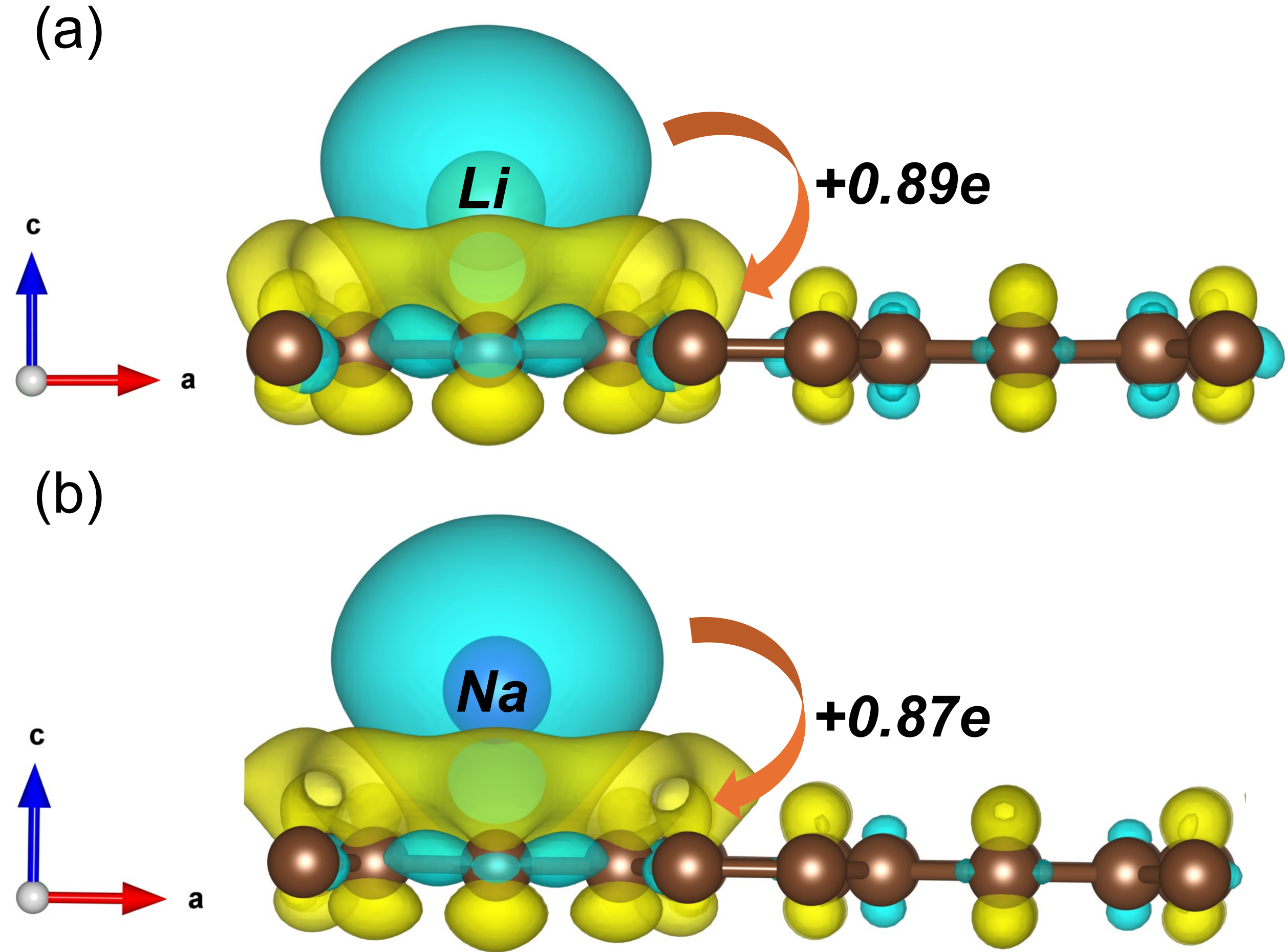}
    \caption{Charge density difference map for single (a) Li and (b) Na adsorption on HOP-graphene structure, where yellow and cyan represent charge accumulation and depletion, respectively.}
    \label{fig:CDD HOP}
\end{figure}

The charge transfer of Li and Na over HOP-graphene was validated by examining the charge density difference (CDD) plot (see Fig.\ref{fig:CDD HOP}, following the relation:
\begin{equation}
\Delta \rho = \rho_{\text{HOP-graphene+Li(Na)}} - \rho_{\text{HOP-graphene}} - \rho_{\text{Li(Na)}}
\end{equation}

\noindent where $\rho_{\mathrm{HOP\text{-}graphene+Li(Na)}}$, $\rho_{\mathrm{HOP\text{-}graphene}}$, and $\rho_{\mathrm{Li(Na)}}$ refer to the electron charge densities of single Li(Na) atoms over HOP-graphene, pristine HOP-graphene, and isolated Li(Na) metal, respectively. In the CDD plot, cyan and yellow regions denote charge accumulation and depletion zones, respectively. One can observe the charge transfer from both Li and Na metals towards the HOP-graphene substrate. As expected, Bader charge analysis confirms a substantial amount of charge that is transferred from Li (+0.89e) and Na (+0.87e) to the carbon anode material.

\subsection{Li and Na diffusion over HOP-graphene}

\begin{figure*} [!ht]
   \centering
    \includegraphics[width=0.75\linewidth]{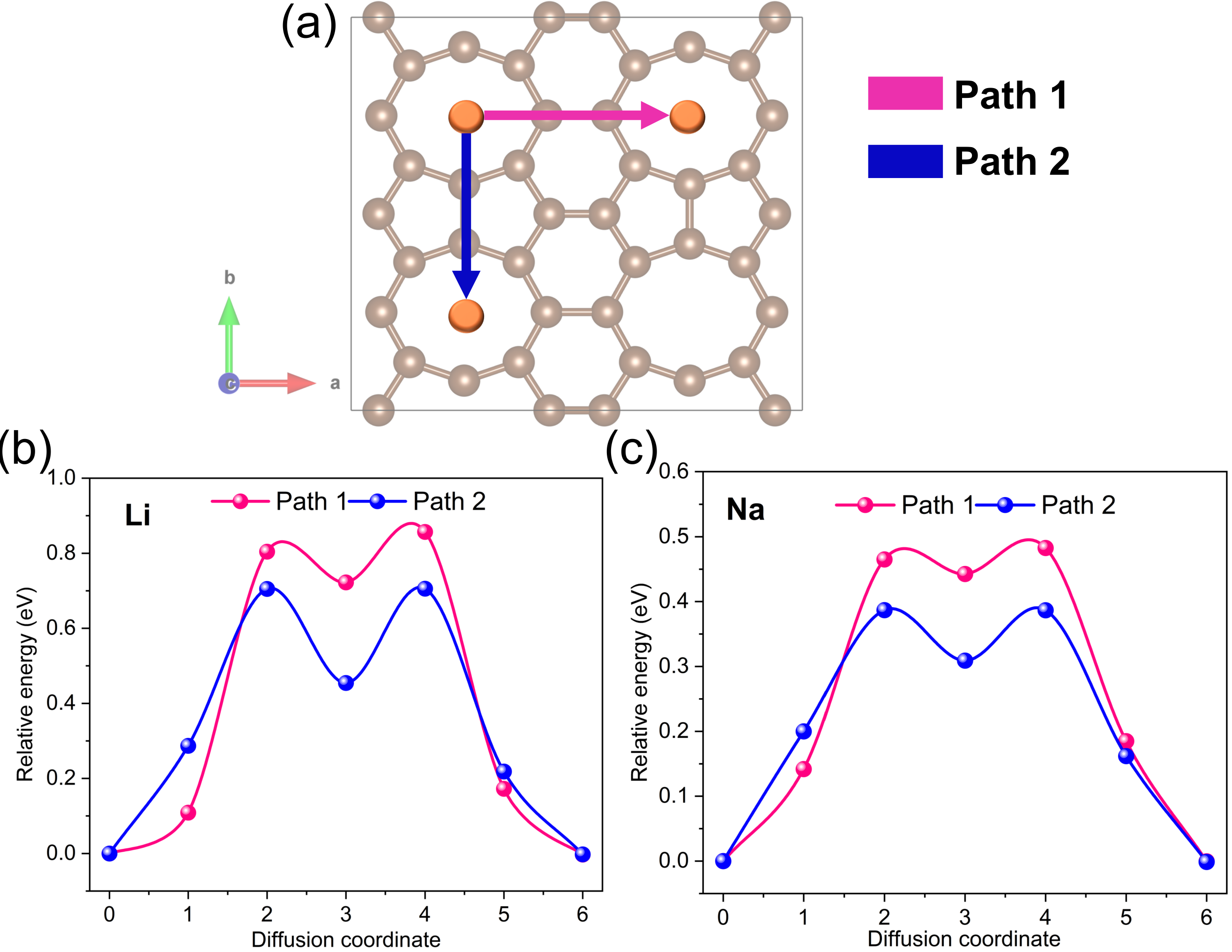}
    \caption{Diffusion pathways and energy barriers for Li and Na migration on HOP-graphene. (a) Top view of the HOP-graphene monolayer showing the two selected diffusion paths: Path 1 (magenta) across hexagonal units and Path 2 (blue) across pentagonal bridges, both connecting adjacent octagonal rings (highlighted by orange spheres). (b) NEB-calculated energy profiles for Li diffusion along Path 1 and Path 2, showing minimum energy barriers of 0.70 eV along Path 2. (c) Corresponding diffusion energy barriers for Na, with the lowest value of 0.39 eV also observed along Path 2.}
    \label{fig:NEB}
\end{figure*}

\begin{figure*} [!ht]
    \centering
    \includegraphics[width=1\linewidth]{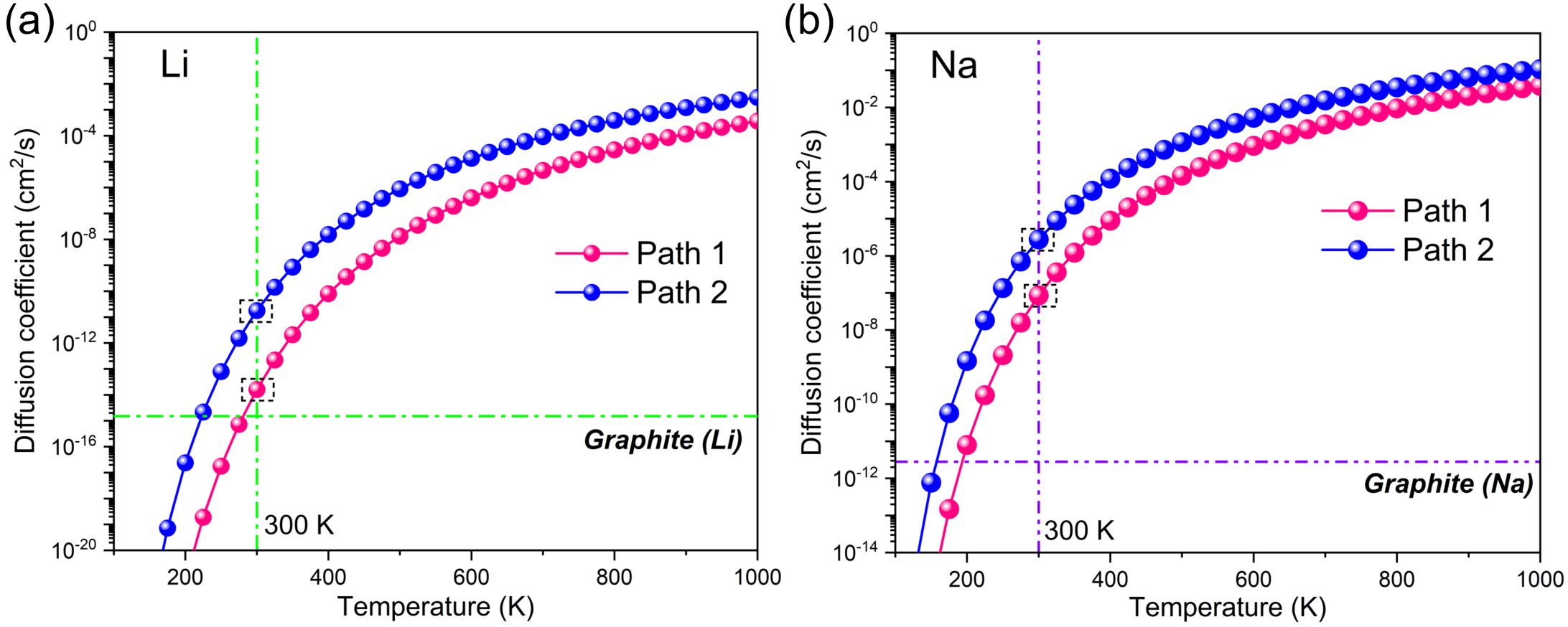}
    \caption{Temperature-dependent diffusion coefficients of Li and Na on HOP-graphene. (a) Diffusion coefficients for Li along Path 1 (magenta) and Path 2 (blue) as a function of temperature, calculated using the Arrhenius equation. (b) Corresponding diffusion behavior for Na. Vertical dashed lines indicate room temperature (300 K), while horizontal dashed lines mark the diffusion coefficients of Li and Na on graphite for comparison.}
    \label{fig:diff coeff}
\end{figure*}

Efficient metal-ion mobility within the electrode is essential for improving the charging and discharging rates of anode materials. In this context, theoretical methods offer valuable insight into the diffusion energy barriers that govern ion transport across the material's surface. To this end, we used the nudged elastic band (NEB) method \cite{jonsson1998nudged,mills1995reversible} to evaluate the migration behavior of Li and Na atoms in HOP-graphene, focusing on their transition from the most stable adsorption sites to neighboring positions. The investigated diffusion pathways are depicted in Fig.~\ref{fig:NEB}(a), where both ions migrate between adjacent octagonal rings: Path 1 proceeds through hexagonal linkages and Path 2 through pentagonal bridges.

The corresponding energy profiles, shown in Figs.\ref{fig:NEB}(b) and \ref{fig:NEB}(c), reveal diffusion barriers of 0.88/0.70 eV for Li and 0.39/0.40 eV for Na along Path 1 and Path 2, respectively. Importantly, the lowest energy barriers, 0.70 eV for Li and 0.39 eV for Na are associated with Path 2, highlighting the role of pentagonal motifs in facilitating ion diffusion. This behavior is consistent with previous theoretical findings\cite{chen2024first, ni2021new}. Moreover, the calculated barriers are comparable to those of other 2D materials, such as biphenylene (0.63 eV for Na)\cite{haouam2024boron}, irida-graphene (0.23 eV for Na)\cite{martins2024irida}, penta-graphyne (0.49/0.61 eV for Li/Na)\cite{deb2022two}, N-graphdiyne (0.80/0.79 eV for Li/Na)\cite{makaremi2018theoretical}, phosphorene (0.76 eV for Li)\cite{carvalho2016phosphorene}, and Petal-graphyne (0.28/0.25 eV for Li/Na)\cite{lima2025petal}, reinforcing the potential of HOP-graphene as a high-performance anode material.

Diffusion coefficients ($D$) were computed to investigate the diffusivity of both Li and Na on HOP-graphene. According to the Arrhenius equation, $D$ can be calculated as follows~\cite{gomez2024tpdh}:
\begin{equation}
D(T) = L^2 \nu_0 \exp\left(-\frac{E_{\text{barr}}}{k_B T}\right),
\label{eq:arrhenius}
\end{equation}

\noindent where $D$ is the diffusion coefficient, $L$ is the migration distance, $\nu_0$ is the vibrational frequency, $k_B$ is the Boltzmann constant ($1.38 \times 10^{-23}$~J/K), and $T$ is the temperature. The vibrational data were obtained from phonon frequencies and used as an empirical vibrational pre-factor, with $\nu_0 = 1 \times 10^{13}$~Hz. The temperature-dependent diffusion coefficients are shown in Figs.~\ref{fig:diff coeff}(a) (Li) and \ref{fig:diff coeff}(b) (Na), considering two estimated migration pathways (Paths 1 and 2). 

At 300 K, Li exhibits diffusion coefficients of $1.62 \times 10^{-14}$ cm$^2$/s along Path 1 and $1.78 \times 10^{-11}$~cm$^2$/s along Path 2, both exceeding the diffusion rate of Li on graphite ($1.50 \times 10^{-15}$~cm$^2$/s) \cite{wang2015diffusion}. In contrast, Na experiences a lower diffusion barrier on HOP-graphene, which leads to significantly enhanced mobility. The corresponding diffusion coefficients at room temperature are $8.57 \times 10^{-8}$ cm$^2$/s for Path 1 and $2.78 \times 10^{-6}$ cm$^2$/s for Path 2, exceeding greatly the value reported for Na on graphite ($2.80 \times 10^{-12}$ cm$^2$/s) \cite{wang2015diffusion}. These results demonstrate that HOP-graphene offers superior ion transport capabilities, which are essential for minimizing electrode degradation and improving the efficiency of charge and discharge processes in battery applications.

\subsection{Storage capacity and OCV profile}

\begin{figure*} [!ht]
    \centering
    \includegraphics[width=0.7\linewidth]{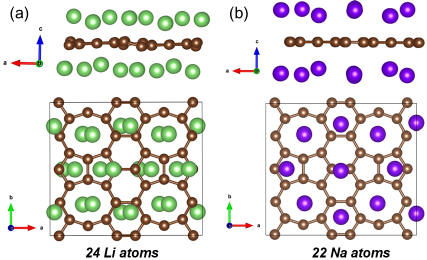}
    \caption{Side and top perspectives for maximum capacity achieved for (a) Li and (b) Na ions over HOP-graphene monolayer.}
    \label{fig:loading}
\end{figure*}

\begin{figure} [!ht]
    \centering
    \includegraphics[width=1\linewidth]{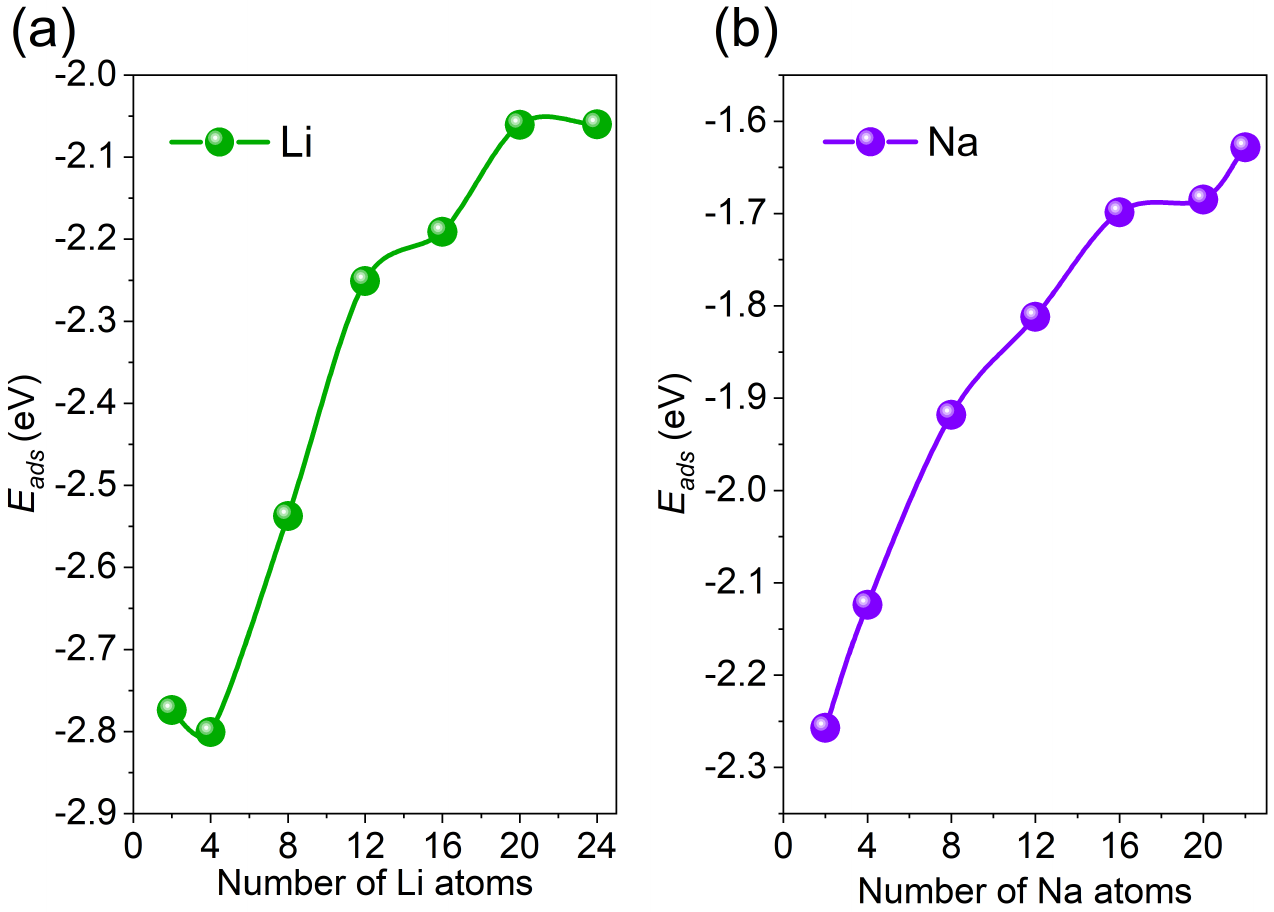}
    \caption{Adsorption energy (E$_\text{ads}$) as a function of number of (a) Li and (b) Na atoms adsorbed into HOP-graphene.}
    \label{fig:ads storage}
\end{figure}

Considering the previous sections, we study Li/Na loading on the HOP-graphene monolayer, beginning with adsorption over the octagonal region. A double-sided symmetrical configuration was employed during the loading process to suppress electrostatic repulsive forces between the alkali metal atoms. To ensure the thermodynamic stability of Li/Na on HOP-graphene, the interatomic distances of Li--Li and Na--Na in their respective bulk phases were monitored as the alkali metal concentration increased. As a result, a maximum loading capacity of 24 Li atoms and 22 Na atoms was achieved. Figure~\ref{fig:ads storage} presents the adsorption energy ($E_\text{ads}$) as a function of Li/Na concentration, with values ranging from $-2.77$ to $-2.06$~eV for Li and from $-2.26$ to $-1.63$~eV for Na. These results confirm that chemical adsorption persists even at the highest metal concentrations, consistent with findings for other 2D materials~\cite{ullah2024theoretical, ding2020assessing, liu2020density}. The side and top views of the fully loaded Li/Na configurations are shown in Fig.~\ref{fig:loading}.

Both alkali metals exhibit a clear tendency to migrate toward the regions previously identified as the most stable, namely the hexagonal and pentagonal patterns. The small atomic radius of Li contributes to a reduced average adsorption height, calculated to be approximately 1.75~\AA. As a result, the symmetric configuration above and below the surface becomes slightly displaced due to the enhanced repulsive interactions associated with this characteristic. In contrast, the simulations suggest a perfectly symmetric sandwich-like structure for Na\@HOP-graphene after full geometric optimization, which is consistent with the larger atomic radius of Na. These observations align with the reported maximum Li/Na coverage behavior in related 2D structures~\cite{peng2023exploring,ma2024novel,zhao2023doping}.

\begin{figure*} [!ht]
    \centering
    \includegraphics[width=0.75\linewidth]{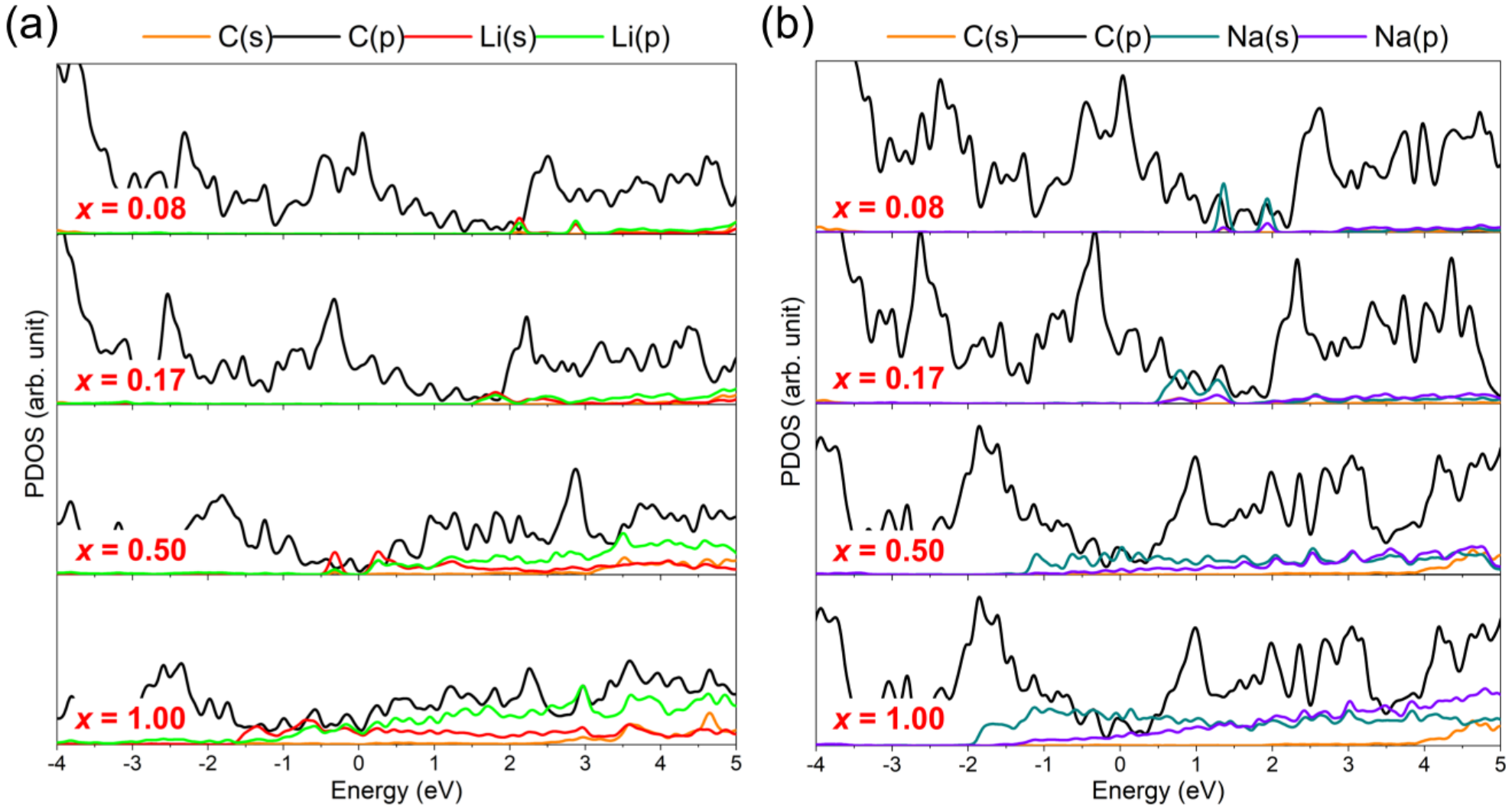}
    \caption{Projected density of states (PDOS) analysis for different concentrations of (a) Li and (b) Na on HOP-graphene structure.}
    \label{fig:PDOS metals}
\end{figure*}

To investigate the electronic response of lithiation and sodiation on the HOP-graphene monolayer, a PDOS analysis was performed, as illustrated in Fig.~\ref{fig:PDOS metals}. This figure presents the PDOS profiles for increasing concentrations of Li (Fig.~\ref{fig:PDOS metals}(a)) and Na (Fig.~\ref{fig:PDOS metals}(b)) adsorbed on the substrate. The metallic nature of HOP-graphene is preserved throughout the entire loading process, which is a key requirement for its application as a practical anode material. Furthermore, introducing additional Li or Na atoms increases the alkali metal states in the conduction band region. For the Li case, there is a pronounced overlap between Li-derived states and the C($p$) orbitals above the Fermi level (set at zero energy). These observations are consistent with the CDD maps obtained for single-atom adsorption (see Fig.~\ref{fig:CDD HOP}), indicating substantial charge transfer from the metallic species to the HOP-graphene surface during Li and Na storage.

\begin{figure*} [!ht]
    \centering
    \includegraphics[width=1\linewidth]{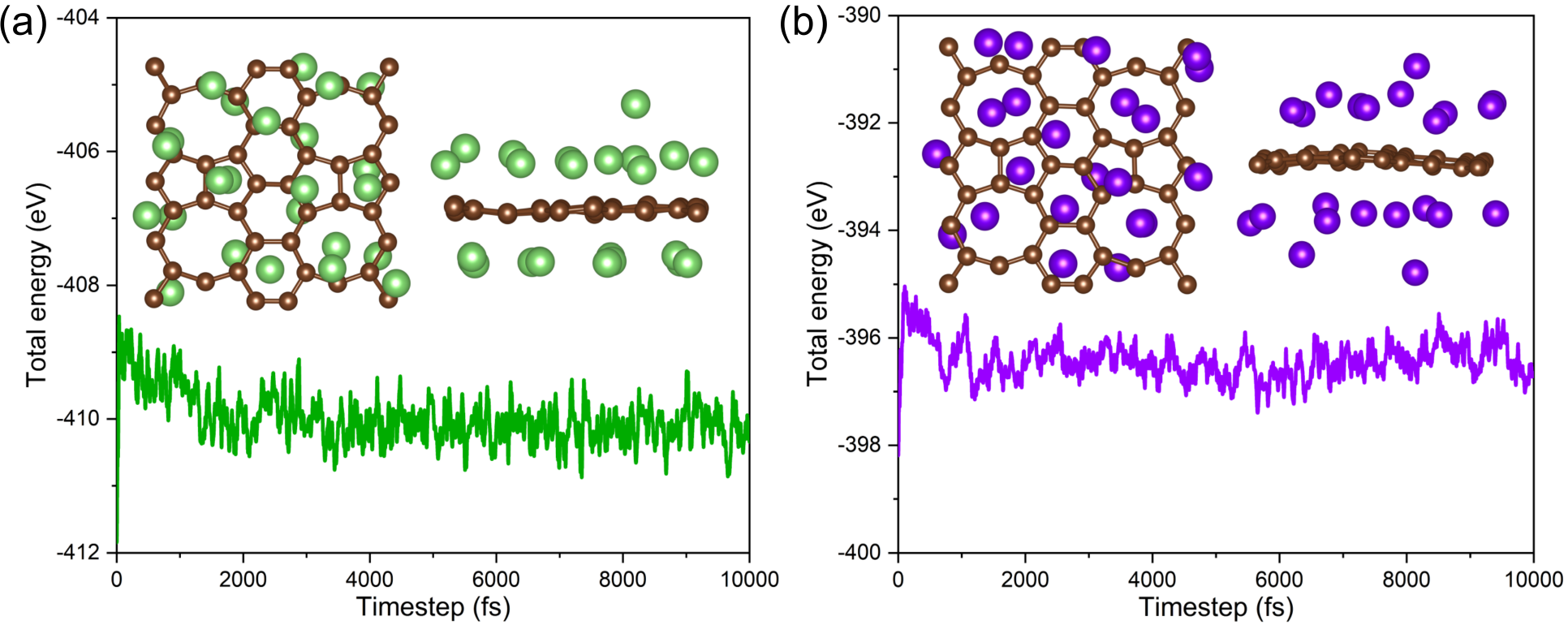}
    \caption{AIMD simulations at 300 K for the fully (a) lithiated and (b) sodiated HOP-graphene.}
    \label{fig:AIMD HOP}
\end{figure*}

AIMD simulations at 300~K were conducted to assess the thermal stability and feasibility of the proposed systems. Figure~\ref{fig:AIMD HOP} shows the total energy fluctuations over a simulation time of 5~ps. For both Li and Na storage systems, only minor energy deviations were observed once the systems reached dynamic equilibrium, remaining below 1.5~eV. Final snapshots reveal that the HOP-graphene substrate remains structurally intact throughout the simulation. Additionally, no aggregation of Li or Na atoms is observed under thermal excitation; instead, the metal atoms continue to interact with the carbon surface. These findings confirm the thermodynamic viability of Li and Na storage on HOP-graphene, with AIMD results supporting their stable behavior and suitability for safe anodic operation.

The theoretical storage capacities of the fully lithiated (Li$_{24}$C$_{40}$) and sodiated (Na$_{24}$C$_{40}$) HOP-graphene structures were estimated using the following equation~\cite{ullah2024theoretical}:
\begin{equation}
C_Q = \frac{nzF}{W} \times 1000, \tag{4}
\end{equation}

\noindent where \( n \) is the number of adsorbed Li/Na atoms, \( z \) is the valence number (equal to 1 for both Li and Na), \( F \) is the Faraday constant (\SI{26801}{Ah/mol}), and \( W \) is the molecular weight of the host material. 

Based on this calculation, the HOP-graphene monolayer exhibits remarkable storage capacities of 1338~mAh/g for Li and 1227~mAh/g for Na, respectively. These values significantly exceed those of conventional graphite anodes for Li/Na (372/35~mAh/g)\cite{armand2008building, wang2016emerging}, highlighting the superior potential of HOP-graphene as an anode material. Furthermore, a comparison with other recently proposed 2D materials confirms the outstanding performance of the HOP-graphene structure. The predicted capacities for Li/Na storage surpass those reported for irida-graphene (1116/1022~mAh/g)~\cite{xiong2024theoretical, martins2024irida}, biphenylene (1302/1075~mAh/g)~\cite{duhan20232, han2022biphenylene}, Petal-graphyne (1004/1004~mAh/g)~\cite{lima2025petal}, 2D-SiC$_7$ (1195/696~mAh/g)~\cite{fatihi2024exploring, yadav2020first}, SiN$_3$ (1146/1146~mAh/g)~\cite{butt2024sin3}, WB$_4$ (708/472~mAh/g)~\cite{masood2024theoretical}, and BeP$_2$ (755/755~mAh/g)~\cite{chen2023metallic}.

\begin{figure*} [!ht]
    \centering
    \includegraphics[width=1\linewidth]{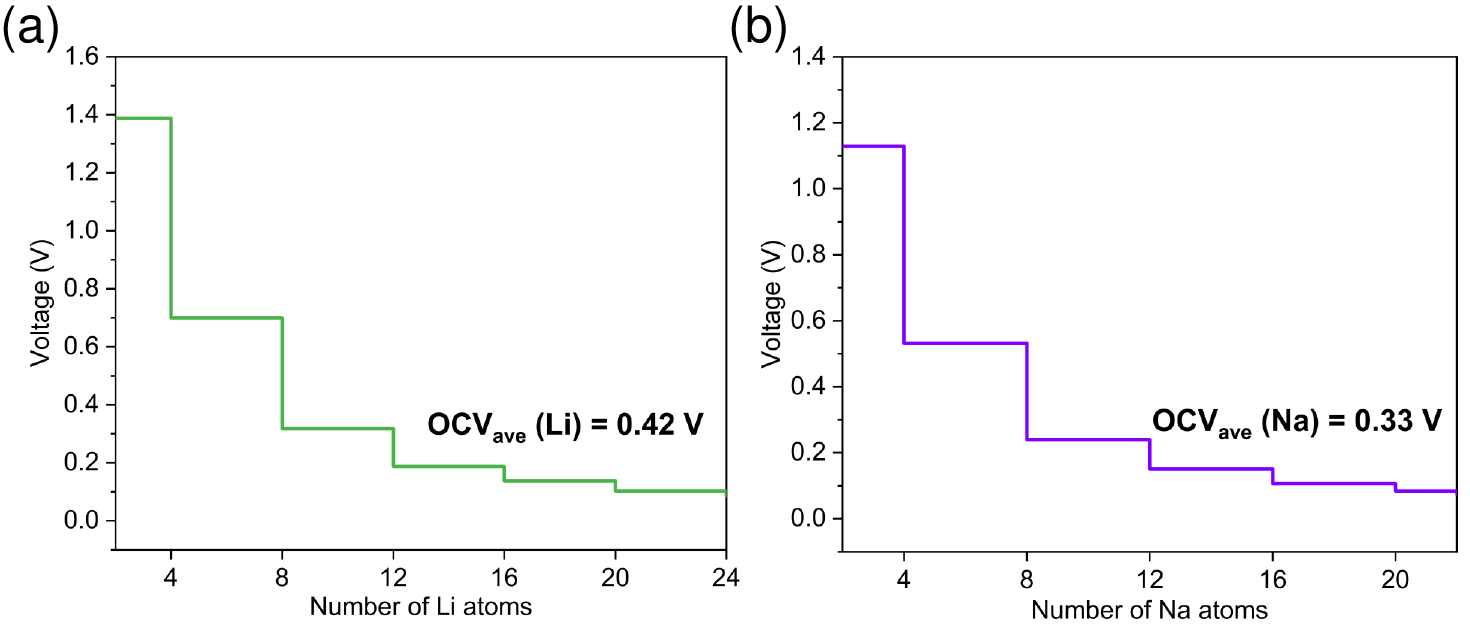}
    \caption{Open-circuit voltage (OCV) profile for (a) Li and (b) Na storage on HOP-graphene monolayer.}
    \label{fig:OCV}
\end{figure*}

The electrode utility of HOP-graphene was assessed by calculating the open-circuit voltage (OCV) profile during lithiation and sodiation. The OCV was estimated based on the energy differences associated with the adsorption of Li/Na atoms~\cite{martins2024irida, surila2025doping}:

\begin{equation}
\text{OCV} = \frac{-(E_{\text{substrate}} + x E_{\text{Li(Na)}} - E_{\text{substrate}+x\text{Li(Na)}})}{x z e}, \tag{5}
\end{equation}

\noindent where $x$ is the number of adsorbed Li or Na atoms, $z$ is the valence (equal to 1 for both Li and Na), and $e$ is the elementary charge. Since negative average adsorption energies indicate thermodynamic favorability, the corresponding positive OCV values suggest stable and safe electrode operation. For practical anode applications, the operating potential should typically remain below 1.0~V~\cite{dua2021twin}. The OCV profiles for Li and Na adsorption on HOP-graphene are shown in Fig.~\ref{fig:OCV}. At the lowest concentrations, the OCV reaches 1.39~V for Li and 1.14~V for Na, gradually decreasing to 0.09~V and 0.07~V, respectively, as the metal coverage increases. The average OCV values are calculated to be 0.42~V for Li and 0.33~V for Na, both of which fall within the optimal range for anodic performance. These results also reflect the electrochemical stability of the system, as negative OCV values can indicate a risk of metal aggregation~\cite{akbar2025dft, shekh2023rational}.

When comparing HOP-graphene with other carbon-based 2D materials, similar average OCV values are reported: TPDH-graphene (0.29~V), biphenylene (0.34~V)~\cite{duhan20232}, irida-graphene (0.32~V)~\cite{martins2024irida}, Pop-graphene (0.45~V)~\cite{wang2018popgraphene}, tetrahexcarbon (0.46~V)~\cite{ma2024novel}, and tolanene (0.25~V)~\cite{ullah2024theoretical}. These similarities further support the potential of HOP-graphene as a high-performance anode material.

\section{Conclusions}

HOP-graphene is a 2D carbon allotrope composed of 5-, 6-, and 8-membered rings, originally predicted through computational simulations. This work comprehensively evaluates its potential as an anode material for lithium- and sodium-ion batteries (LIBs and SIBs) using DFT. Mechanical stability is confirmed via the Born–Huang criteria, while phonon dispersion and AIMD simulations validate its dynamic and thermal robustness. Diffusion analysis reveals that the minimum energy barriers occur between adjacent octagonal rings bridged by pentagonal motifs, with values of 0.70 eV for Li and 0.39 eV for Na. These results suggest favorable ion mobility across the HOP-graphene surface. Corresponding diffusion coefficients are calculated as $1.78 \times 10^{-11}$ cm$^2$/s (Li) and $2.78 \times 10^{-6}$ cm$^2$/s (Na), indicating efficient transport behavior, especially for Na.

Charge transfer analysis using Bader charge and charge density difference (CDD) maps reveals significant electron donation from Li (+0.89e) and Na (+0.87e) atoms to the HOP-graphene substrate. The structure accommodates up to 24 Li and 22 Na atoms adsorbed on both sides, yielding exceptional theoretical storage capacities of 1338 mAh/g and 1227 mAh/g, respectively—surpassing many commercial and predicted 2D anodes. Furthermore, the system exhibits a safe average OCV of 0.42 V (Li) and 0.33 V (Na). These findings position HOP-graphene as a promising 2D carbon-based anode material for next-generation LIB and SIB technologies.

\section*{Data access statement}
Data supporting the results can be accessed by contacting the corresponding author.

\section*{Conflicts of interest}
The authors declare no conflict of interest.

\section*{Acknowledgements}
This work was supported by the Brazilian funding agencies Fundação de Amparo à Pesquisa do Estado de São Paulo - FAPESP (grant no. 2022/03959-6, 2022/14576-0, 2020/01144-0, 2024/05087-1, and 2022/16509-9), and National Council for Scientific, Technological Development - CNPq (grant no. 307213/2021–8). L.A.R.J. acknowledges the financial support from FAP-DF grants $00193.00001808$ $/2022-71$ and $00193-00001857/2023-95$, FAPDF-PRONEM grant $00193.00001247/2021-20$, PDPG-FAPDF-CAPES Centro-Oeste $00193-00000867/2024-94$, and CNPq grants $350176/2022-1$ and $167745/2023-9$. K.A.L.L. acknowledges the Center for Computational Engineering \& Sciences (CCES) at Unicamp for financial support through the FAPESP/CEPID Grant 2013/08293-7. 

\printcredits

\bibliography{cas-refs}

\end{document}